\documentclass[prl,twocolumn,showpacs]{revtex4}
\usepackage{graphicx}
\usepackage[usenames]{color}

\begin{document}

\title{Magnetic quantum tunnelling in $\mathrm{Fe_8}$ with excited nuclei}
\author{$^{1}$Oren Shafir, $^{1}$Amit Keren}
\author{$^{2}$Satoru Maegawa, $^{2}$Miki Ueda }
\author{$^{3}$Efrat Shimshoni}
\affiliation{$^{1}$Department of Physics, Technion-Israel Institute of Technology, Haifa
32000, Israel}
\affiliation{$^{2}$Graduate School of Human and Environmental Studies, Kyoto University,
Kyoto 606-8501, Japan}
\affiliation{$^{3}$Department of Physics, Bar-Ilan University, Ramat Gan 52900, Israel}

\date{\today}

\begin{abstract}
We investigate the effect of dynamic nuclear spin fluctuation on quantum
tunneling of the magnetization (QTM) in the molecular magnet Fe$_{8}$ by
increasing the nuclei temperature using radio frequency (RF) pulses before
the hysteresis loop measurements. The RF pulses do not change the electrons
spin temperature. Independently we show that the nuclear spin-spin
relaxation time $T_{2}$ has strong temperature dependence. Nevertheless, we
found no effect of the nuclear spin temperature on the tunneling
probability. This suggests that in our experimental conditions only the
hyperfine field strength is relevant for QTM. We demonstrate theoretically
how this can occur.
\end{abstract}

\maketitle
\affiliation{Department of Physics, Technion-Israel Institute of Technology, Haifa
32000, Israel}
\affiliation{Graduate School of Human and Environmental Studies, Kyoto University,
Kyoto 606-8501, Japan}
\affiliation{Department of Physics, Bar-Ilan University, Ramat Gan 52900, Israel}

The importance of nuclei to quantum tunnelling of the magnetization (QTM) in
Fe$_{8}$, was demonstrated experimentally by Wernsdorfer \textit{et al.}~%
\cite{Wernsdorfer,Sessoli}. They compared the tunnel splitting $\Delta $ of
the standard Fe$_{8}$ sample with a deuterated sample and a sample where $%
^{56}$Fe was replaced partially with $^{57}$Fe. The enrichment with
deuterium causes a decrease of $\Delta $, in accord with the decreased
hyperfine field (HF) \cite{Wernsdorfer2}. Similar conclusion was obtained
with the $^{57}$Fe enrichment. However, the exchange of isotopes does not
only vary the strength of the HF exerted on the molecule: it also changes
the nuclear spin-spin relaxation rate $T_{2}$. Both quantities might be
important for the nuclear-assisted tunnelling process \cite{ProkofevPRL98}.
Isotope substitution can not tell if only one or both quantities are
relevant. Therefore, it is not yet established experimentally how exactly
nuclei impact the tunnelling process.

The experiment reported here aims at distinguishing between the contribution
of the HF and $T_{2}$. We present magnetization measurements of Fe$_{8}$
during field sweep after transmitting radio frequency (RF) at the protons
resonance. This transmission raises the protons temperature without changing
the electrons temperature due to the enormous proton spin-lattice relaxation
time $T_{1}$ which is longer than $1000$~sec at sub-Kelvin temperatures \cite%
{UedaPRB02}. We also present proton spin-spin relaxation rate $T_{2}$
measurements and show that $T_{2}$ varies with temperarature. When internal
field fluctuations are slow, as in our case, $T_{2}$ is a property internal
to the nuclear spin system. Therefore, we argue that razing the nuclei
temperature must lead to $T_{2}$ variations with no modification to the
hyperfine field. Our major finding is that $T_{2}$ is not a relevant
parameter for QTM.

\begin{figure}[tbp]
\includegraphics[width=2.7in]{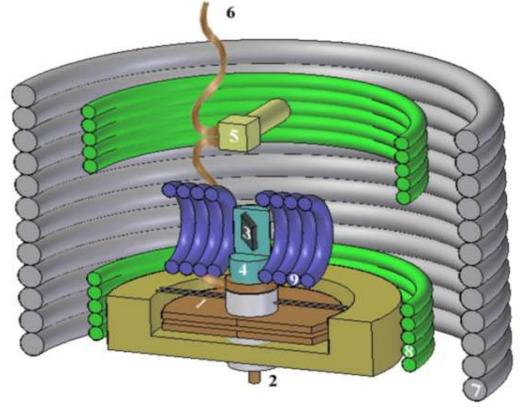} {}
\caption{(Color online) Cross sectional
view of the Faraday balance with: (1) movable plate of the capacitor, (2)
screw for capacitor's fixed plate height adjustment, (3) sample, (4) PCTFE,
(5) gold plated casing of the thermometer, (6) thermal link to the DR mixing
chamber, (7) main coil, (8) gradient coils, (9) RF coil. }
\label{setup}
\end{figure}

For this experiment, a Faraday force magnetometer shown in Fig.~\ref{setup}
was constructed inside the inner vacuum chamber of a dilution refrigerator
(DR) following the design of Sakakibara \textit{et al.\ }\cite{Sakakibara},
with the addition of an RF coil. This magnetometer is suitable for
measurements in high fields and at sub-Kelvin temperatures with no metallic
parts near the sample. This is important for minimizing warming metallic
parts with the RF. The DR is equipped with a main superconducting magnet
that produces the field $H$, and two oppositely wound superconducting
magnets that produce a field gradient.

The sample is mounted on the small load-sensing device made of two parallel
plates variable capacitor. The movable plate is suspended by two pairs of
orthogonal crossed 0.2 mm diameter phosphor bronze wires attached to it with
epoxy. The static lower plate was mounted on an epoxy screw, for adjusting
the initial capacity $C_{0}$. When the sample is subjected to a spatially
varying magnetic field $B$, it will experience a force $\mathbf{F}%
=M_{z}(\partial B_{z}/\partial z)\hat{z}$. This force is balanced by the
wires. The displacement of the plate is proportional to $F$ and can be
detected as a capacitance C change. The total capacitance response is then
given by 
\[
C_{0}^{-1}-C^{-1}=a\cdot M_{z}(\partial B_{z}/\partial z) 
\]
where $a$ is a constant that depends on the elastic properties of the wires.

The sample is grown by the method described in Ref ~\cite{Weighardt} and is
20~mm$^{3}$. It is oriented with its easy axis parallel to the magnetic
field $H$. The sample is glued with GE-varnish to
Poly-Chloro-TriFluoro-Ethylene (PCTFE), a fluorocarbon-based polymer, which
has no hydrogen and is suitable for cryogenic applications. The bottom of
the PCTFE is connected by a thermal link to the DR mixing chamber which
produces the cooling, and to the movable plate. Approximately 2 cm above the
sample, on the thermal link, there is a calibrated thermometer (RuO$_{2}$
R2200) in a gold plated casing. It is important to mention that the sample
is in vacuum with no exchange gas, and therefore its temperature $T$ is not
exactly the same as the temperature of the thermometer. However, this is not
a problem in our experiment since below 400~mK the magnetization jumps of Fe$%
_{8}$ are temperature-independent~\cite{Sangregorio}.

\begin{figure}[tbp]
\includegraphics[width=2.5in]{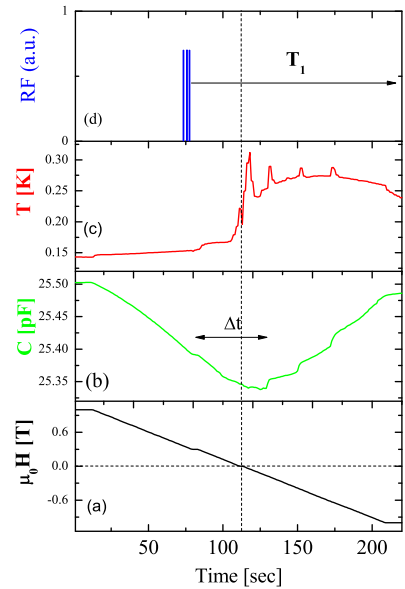} {}
\caption{(Color online) The scheme of the
measurements showing: the magnetic field sweept from positive to negative
(a), the capacitance (b), the temperature (c), and the RF transmission (d). $%
\Delta t$ is the time from the transmission to the first capacitance
(magnetization) jump. $T_{1}$ is the nuclear spin-lattice relaxation time. }
\label{cwithrf}
\end{figure}

In the magnetization experiments we apply a field of +1 T and wait until
thermal equilibrium is reached. We then record the field value [Fig.~\ref%
{cwithrf}(a)], capacitance [Fig.~\ref{cwithrf}(b)], and temperature [Fig.~%
\ref{cwithrf}(c)] as the field is swept from $+1$~T to $-1$~T at a rate $%
dH/dt=0.5$ T/min. While we sweep the magnetic field from positive to
negative, we stop for several seconds at$0.3$~T ($12.71$ MHz) where we
transmit the RF in the form of pulses as shown in Fig.~\ref{cwithrf}(d). All
attempts to deliver RF at a negative field resulted in immediate
magnetization jumps, hence, the choice to transmit at a positive field.
During the transmission, the temperature rises by $20$~mK. When the field
changes sign there is a larger temperature increase of $150$ mK due to eddy
currents. None of these temperature changes are enough to generate
magnetization changes. To make the measurement with and without RF as
similar as possible, we stopped at $0.3$~T for several seconds even when we
do not transmit RF.

We first concentrate on the capacitance versus time, for a full sweep shown
in Fig.~\ref{cwithrf}(b). At times where the field is positive the
capacitance is a smooth function of time (and field). This is because the
spins are at their ground state for all positive fields and have nowhere to
tunnel to. Once the field becomes negative, clear jumps in the capacitance
are observed, indicating jumps in the magnetization that are taking place
when tunnelling occurs between molecular spin states. The time it takes to
sweep from the end of the RF transmission to the first jump is $\Delta t=60$
sec. This time is much shorter than the nuclear $T_{1}$, as demonstrated in
Fig.~\ref{cwithrf}(d). Therefore, the nuclei are expected to be excited when
the Fe$_{8}$ spins are tunnelling. We avoided RF transmission at fields
higher than $0.3$~T, and used high sweep rate in order to keep $\Delta t$
short. Finally, Fig.~\ref{cwithrf}(c) shows that magnetization jumps are
accompanied by temperature spikes. These are discussed in a separate paper 
\cite{Shafir}.

\begin{figure}[tbp]
\includegraphics[width=2.4in]{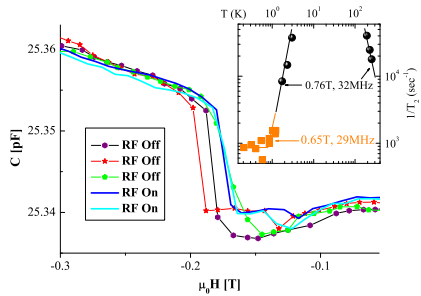}
\caption{(Color online) Capacitance
measurements as a function of field swept from positive to negative, with
and without RF. The inset shows the temperature dependence of $1/T_{2}$ in Fe%
$_{8}$ measured near the free proton resonance condition. }
\label{rfjumps}
\end{figure}

The results of measurements with and without the RF are summarized in Fig.~%
\ref{rfjumps}. We focus on the first magnetization jump which is closest to
the time of RF irradiation. The solid lines show sweeps with RF and the
solid with symbols lines are sweeps without RF. We repeated these runs
several times and found that within our experimental resolution, and
stability between individual sweeps, no effect of the RF can be detected.

To appreciate this result we performed $T_{2}$ measurements inside the
mixing chamber of a DR and He cryostat using a more standard NMR setup and
coil at fields of 0.76~T and 0.65~T and frequencies of $32$~MHz and $29$~MHz
respectivly. At these conditions the resonance field for most of the protons
is not shifted from the free proton resonance, and the line width $\Delta H$
is on the order of $200$ mT \cite{yamasaki}. The results of $1/T_{2}$ are
presented in the inset of Fig.~\ref{rfjumps}. $T_{2}$ varies from less than $%
10^{-4}$~sec at $T=3$~K to $10^{-3}$ sec below $T=0.5$~K. Between $4$~K and $%
150$~K $T_{2}$ is so short that no signal could be detected. Finally, due to
the huge time scale difference between $T_{2}$ and $T_{1}$ at all
temperatures, it is reasonable to assume that $T_{2}$ is determined by
nuclear spin-spin coupling only.

\begin{figure}[tbp]
\includegraphics[width=3.2093in]{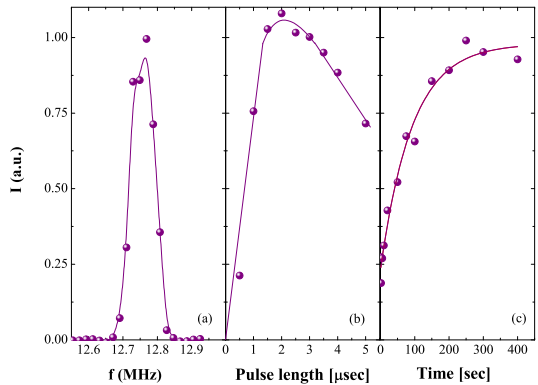}
\caption{(Color online) Echo intensity
at 140~mK from the GE-varnish gluing the sample as a function of field (in
frequency units) (a), pulse length (b), and time after saturation (c). The
solid lines are guides to the eye. }
\label{nmr}
\end{figure}

In the set up with both RF and magnetization shown in Fig.~\ref{setup} it is
difficult to detect the proton signal due to the poor filling factor in the
Helmholtz coil, the broad line width, and the extremely long $T_{1}$.
However, the GE-varnish gluing the sample has relatively narrow line and
shorter $T_{1}$. We therefore use the varnish signal at $T=140~$mK to
confirm the delivery of the RF radiation to the sample, to measure the
strength of the RF field $H_{1}$, and to test our ability to saturate the
nuclear transitions. First, we measured the echo intensity as a function of
applied field at constant frequency of $12.71$~MHz (0.3~T) using a $\pi
/2-\pi $ pulse sequence. As shown in Fig.~\ref{nmr}(a), the full width at
half maximum is only $4\pm 0.5$ mT. A similar line width was found for the
varnish in our standard NMR spectrometer at $5$ K. This ensures that we
deliver the radiation to the center of the RF coil. Second, we determined
the optimal pulse length. The echo intensity as a function of the pulse
length $t_{\pi /2}$ is presented in Fig.~\ref{nmr}(b). The maximum echo
intensity was found at $t_{\pi /2}$ =1.5 $\pm $ 0.5 
$\mu$%
sec. From $\gamma H_{1}t_{\pi }=\pi /2$ we calculated $H_{1}$ to be $24\pm 4$
mT. Finally, we determined $T_{1}$, as presented in Fig.~\ref{nmr}(c) by
saturating the proton transitions with a train of pulses, and then measuring
the recovery of the signal at a time $t$ using $\pi /2-\pi $ pulses. The
pulse train equilibrates the up and down proton spins population. We found
that the GE-varnish $T_{1}$ is only $100$ sec. More importantly, Fig.~\ref%
{nmr}(c) demonstrates our ability to saturate the proton transitions.

The above measurement allows us to estimate the variation in the nuclear $%
T_{2}$ at the time electronic spins are tunneling due to our RF irradiation.
First, we examine how many protons we excite. Since our $H_{1}$ is smaller
than the Fe$_{8}$ line width $\Delta H$, our direct pulses excite only $%
H_{1}/\Delta H=10\%$ of the total number of protons. However during the
transmission and after it, spin diffusion is taking place spreading the
nuclear temperature among all nuclei. The diffusion coefficient $D$ is given
by $D=Wr^{2}$ where $W$ is flip-flop rate of neighboring nuclei, and $r$ is
the distance between them \cite{Abragam}. For dipolar coupling $W>(\gamma
^{2}\hbar /r^{3})^{2}/(\gamma \Delta H)$ where $\gamma ^{2}\hbar /r^{3}$ is
the strength of the dipolar interaction, and $1/\gamma \Delta H$ is a lower
limit on the density of states \cite{Abragam}. The time it takes for the
heat to spred among all nuclei in a unite cell of volume $V$ is $V^{2/3}/D$,
which is less than $10$ sec. Therefore, all nuclei should be warm before the
first tunnelling event is taking place.

Second, we evaluate by how much the irradiate nuclei cool during the time
between transmission and tunnelling. For this we employ the equation 
\[
\frac{1}{T{_{e}}}(1-\exp (-\Delta t/T_{1}))=\frac{1}{T{_{n}}} 
\]%
where $T_{e}$ is the electron's temperature (140 mK) and $T_{n}$ is the
nuclei temperature. Immediately after the RF pulses ($\Delta t=0$) $%
T_{n}=\infty $. As $\Delta t$ grows $T_{n}$ decreases until at $\Delta
t=\infty $ it reaches the electrons temperature again. This equation
suggests that the nuclei temperature at the time of the tunnelling is well
above $3$~K where $T_{2}$ increases by a factor $14$ from its value at $140$%
~mK. Therefore, we conclude that changing $T_{2}$ by an order of magnitude
has no effect on the tunnelling probability, for our sweep rate.
 
To demonstrate that it is conceivable to have an isotope effect, yet no
dependence on $T_{2}$, we analyze an effective model for the dynamics of the
system in the vicinity of a resonant transition between molecular spin
levels $m$ and $m^{\prime }$. This is essentially the Landau-Zener (LZ)
problem with the addition of a transverse magnetic noise. The effective
Hamiltonian, describing a spin $1/2$ with a resonance tunnel splitting $%
\Delta $, a time-dependent magnetic field in the $z$ direction, and a
fluctuating magnetic field in the $x$ direction, is given by 
\begin{equation}
\ \mathcal{H}=\alpha tS_{z}+\Delta S_{x}+B_{x}^{e}(t)S_{x}\;.
\label{Schrodinger}
\end{equation}%
Here $S_{x}=\frac{1}{2}\sigma _{x}$ and $S_{z}=\frac{1}{2}\sigma _{z}$,
where $\sigma _{x}$ and $\sigma _{z}$ are the Pauli matrices, and $\alpha
=2g\mu _{B}\left( {m-m^{\prime }}\right) dH/dt$. We assume that the
stochastic field $B_{x}^{e}(t)$ has a correlation function 
\begin{equation}
\ \left\langle B_{x}^{e}{(t)}B_{x}^{e}{(t}^{\prime }{)}\right\rangle
=\left\langle B_{x}^{e}{}^{2}\right\rangle \exp (-\left\vert t-t^{\prime
}\right\vert /\tau _{c})  \label{Correlation}
\end{equation}%
where $\tau _{c}$ is the correlation time. $B_{x}^{e}$ is related to the
hyperfine field (see below), and ${\tau _{c}}$ stands for $T_{2}$. We
consider only a transverse fluctuating field since for the $-10$ to $9$
transition, the measured tunnel splitting $\Delta =3\times 10^{-7}$ K \cite%
{delta}, and our sweep rate, the sudden limit is obeyed, namely, $\Delta /%
\sqrt{\hbar \alpha }\ll 1$. In this case it is well established that a
stochastic field coupled to the $z$ direction of the spin has no effect on
the LZ tunnelling probability \cite{ZTheory}.

We next write the wave function as $\Psi (t)=\widetilde{C}_{-}(t)\left\vert
-\right\rangle +\widetilde{C}_{+}(t)\left\vert +\right\rangle $, where $%
\left\vert \pm \right\rangle $ denote eigenstates of $\sigma _{z}$. Defining 
$C_{\pm }(t)=\exp (\pm i\alpha t^{2}/4\hbar )\widetilde{C}_{\pm }(t)$ and
introducing a dimensionless time variable $y=t\sqrt{\alpha /\hbar }$, the
Schr\"{o}dinger equation can be expressed in the integral form 
\begin{eqnarray}
C_{\mp }(\infty ) &=&\frac{i}{2\sqrt{\alpha \hbar }}\int_{-\infty }^{\infty
}\left( \Delta +B_{x}^{e}(y)\right)  \label{Solution} \\
&&e^{\pm iy^{2}/2}C_{\pm }(y)dy.  \nonumber
\end{eqnarray}%
Assuming the initial conditions $C_{+}(-\infty )=1$ and $C_{-}(-\infty )=0$,
the tunnelling probability is given by $P=1-\left\langle \left\vert
C_{-}(\infty )\right\vert ^{2}\right\rangle $ where $\left\langle
{}\right\rangle $ stands for an average of stochastic field realizations. In
the sudden limit, $C_{+}$ does not change much. If, in addition, $B_{x}\ll 
\sqrt{\alpha \hbar }$ then $C_{+}$ under the integral can be replaced by 1
to first order in $\Delta $ and $B_{x}$. This yields 
\begin{equation}
\ P = 1 - \frac{1}{{4\alpha \hbar }}\int\limits_{}^\infty  {\int\limits_{ - \infty }^{} {dxdy\left( {\Delta ^2  + \left\langle {B_x^{e2} } \right\rangle e^{\nu \left| {x - y} \right|} } \right)} } e^{i\left( {y^2  - x^2 } \right)/2} 
\end{equation}%
where we have used Eq.~\ref{Correlation} with $\nu \equiv 1/\tau _{c}\sqrt{%
\alpha /\hbar }$. This yields 
\begin{equation}
\ P=1-\pi (\Delta ^{2}+\left\langle B_{x}^{e}{}^{2}\right\rangle )/2\hbar
\alpha ,  \label{Final}
\end{equation}%
in which there is \emph{no dependence} on the parameter $\nu $. The
transition probability is therefore dependent on the HF strength, but not on
its correlation time $\tau _{c}$.

To make this derivation attractive $\left\langle
B_{x}^{e}{}^{2}\right\rangle ^{1/2}$ must be of the order of the measured
tunnel splitting. When converting the Fe$_{8}$ problem to the two level LZ
problem $B_{x}^{e}$ is scaled down from the field $B_{x}$ the nuclei
produce, since $B_{x}$ has a matrix element between the $m$ and $m^{\prime }$
states only in the $\left\vert {m-m^{\prime }}\right\vert $'th order of
perturbation theory. As a consequence, Garanin and Chudnovsky \cite%
{GaraninPRB97} showed that 
\begin{equation}
B_{x}^{e}=\frac{2D}{(m^{\prime }-m-1)!^{2}}\sqrt{\frac{(S+m^{\prime })!(S-m)!%
}{(S-m^{\prime })!(S+m)!}}\left( \frac{B_{x}}{2D}\right) ^{m^{\prime }-m}
\end{equation}%
where $D=0.27$~K is the Fe$_{8}$ single ion anisotropy coefficient. Protons
produce a field of the order of $1-10$~mT inside a solid, corresponding to $%
B_{x}$ of $0.01-0.001$~K, which is not small. However, in our case $%
m^{\prime }-m=19$ therefore $B_{x}^{e}$ is practically zero. For $B_{x}^{e}$
to be relevant there has to be a shortcut in the tunneling process that will
make $m^{\prime }-m$ smaller and of the order of $6-7$. It is reasonable
that such a shortcut exists since we, and other researchers \cite{delta},
see only four magnetization jumps and not $10$.

To summarize, we exploit the strong temperature dependence of the nuclear
spin-spin relaxation time $T_{2}$ around 1~K in order to test the effect of
nuclear fluctuations on quantum tunneling of the magnetization. Since in our
case $T_{2}$ is a property internal to the nuclear spin system, we change it
by warming only this system with radio frequency transmitted at the protons
resonance. We then measure the size of the magnetization jumps due to
tunneling. During the magnetization measurements the nuclei stay warm due to
the enormously long spin-lattice relaxation time $T_{1}$. We found no effect
of the nuclear spin temperature on the magnetization jump and argue that $%
T_{2}$ is a parameter irrelevant to the tunneling probability in our
experimental conditions. We present a calculation demonstrating that nuclear
spins can, indeed, affect the tunneling via their hyperfine field strength
only.

We are grateful for RBNI Nevet program and Israeli ministry of science
"Tashtiot" program for supporting this research.

\end{document}